\documentclass[preprint,psfig]{aastex}
\usepackage{epsfig}
\usepackage{psfig}

\def\edcomment#1{\iffalse\marginpar{\raggedright\sl#1\/}\else\relax\fi}
\reversemarginpar

\def\beq{\begin{equation}}
\def\eeq{\end{equation}}

\begin{document}
\title{New Statistical Methods for Analysis of Large Surveys: Distributions and 
Correlations} 
\author{Vahe' Petrosian}
\affil{Center for Space Science and Astrophysics, Varian 302c, Stanford 
University, Stanford, CA 94305-4060}

\begin{abstract}
The aim of this paper is to describe new statistical methods for determination
of the correlations among and distributions of physical parameters from a
multivariate data with general and arbitrary truncations and selection biases.
These methods, developed in collaboration with B.  Efron of Department of
Statistics at Stanford, can be used for analysis of combined data from many
surveys with different and varied observational selection criteria.  For clarity
I will use the luminosity function of AGNs and its evolution to demonstrate the
methods.  I will first describe the general features of data
truncation and present a brief review of past methods of analysis.  Then I 
will
describe the new methods and results from simulations testing their accuracy.
Finally I will present the results from application of the methods to a sample
of quasars.
\end{abstract}

\section{INTRODUCTION}

One of the important ways of testing the models of AGN, or any other 
astrophysical source, is through the investigations of the distributions, 
ranges, and more importantly the correlations among, the relevant physical 
characteristics, such as luminosity, spectrum, redshifts or distances. A 
reliable determination of these features requires large samples. As evident 
from papers presented in this proceedings the samples are becoming larger and 
larger. Combining the samples, however, is a very challenging task, because 
different samples are obtained by different instruments and techniques, so that 
they suffer from different and varied selection biases and data truncations. 
Overcoming these biases requires care.

The primary goal of this paper is to describe some of the relatively new 
methods we have developed  at Stanford over the past decade (Efron \& Petrosian 
1992, 1999). These methods are very general and are applicable to any data with 
well defined but arbitrary truncations. We have applied these to various 
astrophysical data such as solar flares (Lee, Petrosian, \& McTiernan 1993, 
1995), gamma-ray burst (see e.g. Lloyd, Petrosian, \& Mallozzi 2000) and 
quasars (Meloney \& Petrosian 1999). Instead of using abstract mathematical 
symbols, the method will be 
demonstrated using the luminosity function of AGNs and its 
cosmological evolution, {\it i.e.} its variation with redshift $z; \Psi(L,z)$.
Without loss of generality, we can write the luminosity function as 

\beq\label{lf}
\Psi(L,z) = \rho (z) \psi(L/g(z),\alpha_i) / g(z),
\eeq
where $\rho(z)$ describes the co-moving density evolution and $g(z)$ (with
$g(0) = 1$) describes the luminosity evolution of the population with $L_o =
L/g(z)$ as the luminosity adjusted to its present epoch value; $\psi
(L_o, \alpha_i)$ gives the local luminosity function.  Here I explicitly
include the shape parameters $\alpha_i$, which could also vary with redshift. A
surprising result has been the absence of evidence for a strong shape evolution. 
In this paper I ignore such effects and concentrate on the 
determination of the the density and luminosity evolution functions $\rho(z)$
and $g(z)$.

In the next section we describe various kind of truncations and in \S 3 give a
brief summary of some of the past methods used for this kind of analysis. 
The new methods and their accuracy are described in \S 4 and a sample of
results from application to AGNs are summarized in \S 5. 

\section{TYPES OF TRUNCATIONS}

The left panel of Figure 1 shows a set of arbitrary data points labeled as
luminosity $L$ and redshift $z$ and several generic truncations.  The 
distribution
may be truncated parallel to the axis as shown by the dotted lines.  This only
limits the observed ranges of the variables but does not introduce any bias.  
This
kind of data will be referred to as untruncated data.  However, this is rarely 
the
case for astronomical data, and in general, one is dealing with cases where the
truncation is not parallel to the axis.  The simplest and most common case is 
when
the data suffers a one-sided truncation from below as shown by the solid curve.
This is the case for magnitude or flux limited samples; $L\geq 4\pi
d_L^2(\Omega_i, z) f_{min}$, where $f_{min}$ is the limiting flux and
$d_L(\Omega_i, z)$ is the luminosity distance at $z$ for an assumed cosmological
model represented by $\Omega_i$.  In some cases the data may be truncated from
above as shown by the dashed curve.  The statistical methods do not distinguish
between truncation from above or below.  However, the data analysis is affected
when there are truncations both from {\bf above and below}.  This is the case 
for
some AGN samples.  The situation becomes even more complex when the truncation
boundaries are not monotonic, or when one tries to combine samples with 
different
truncations, say different upper and lower flux limits.  Such variation may be
present even within a given catalog where data taken at different times and
directions in the sky may have different limits.  The most general truncation is
when each data point, say [$L_i, z_i$], has its individual upper and lower 
limits,
$L_i^-<L_i<L_i^+$ and $z_i^-<z_i<z_i^+$, as shown by the large cross for one 
point
on Figure 1.  The methods we have developed can treat this most general
truncation.

\begin{figure}[ht]
\leavevmode\centering{
\psfig{file=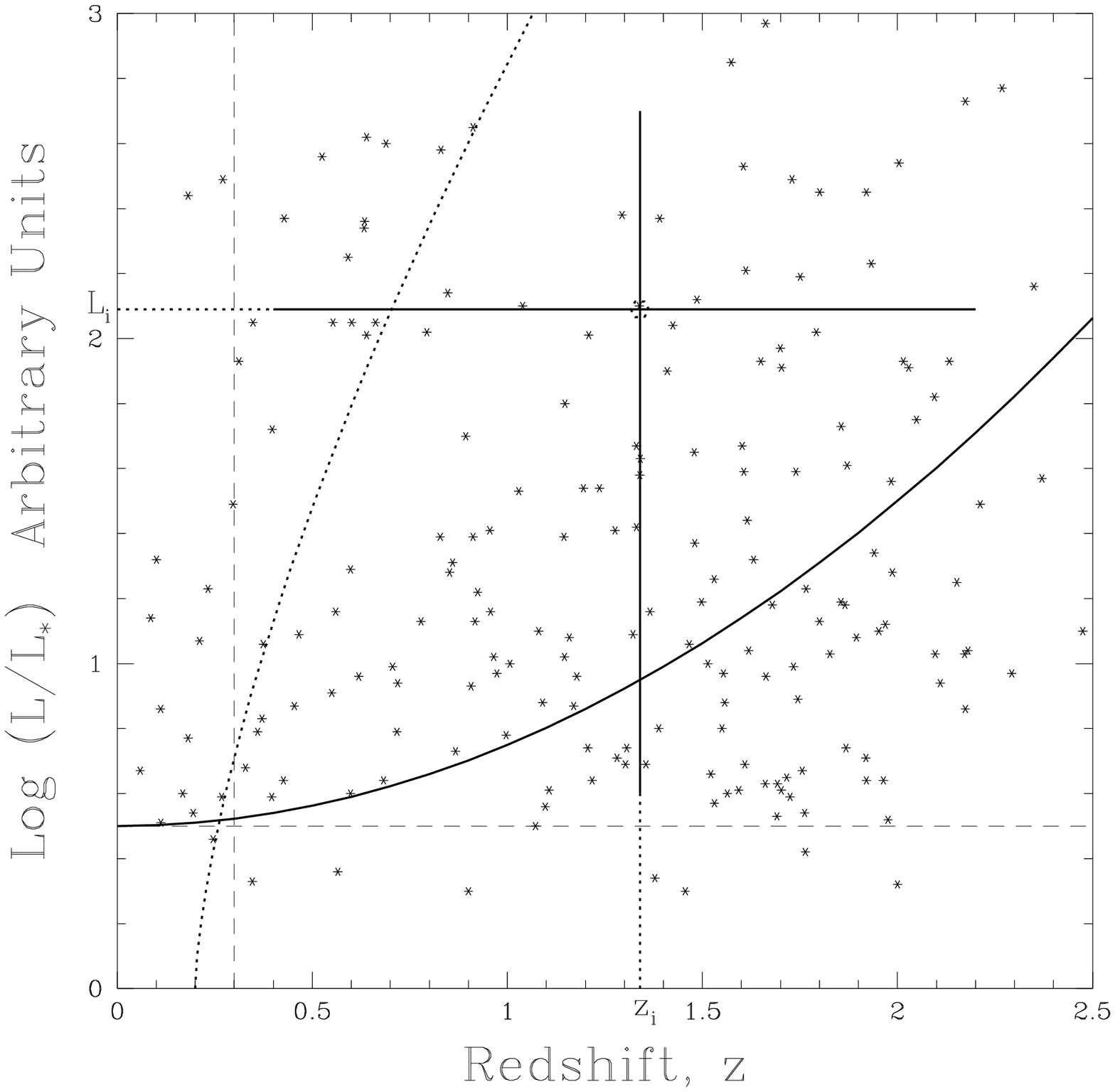,width=0.48\textwidth,height=0.48\textwidth}
\psfig{file=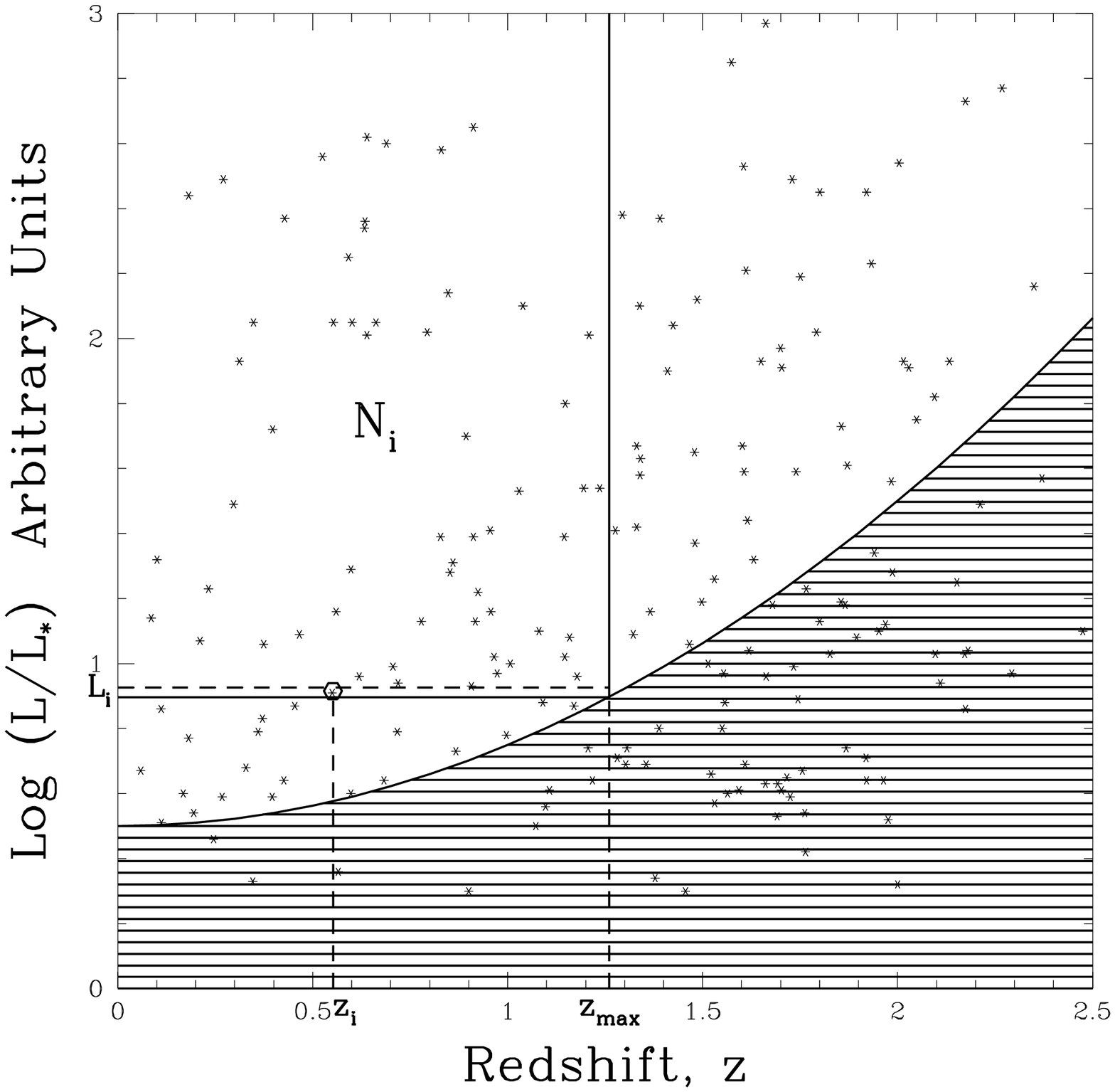,width=0.48\textwidth,height=0.48\textwidth}}
\caption{
{\bf  Left Panel:\,} Demonstration of various types of data 
truncations: Parallel to axis (dotted lines), from below (the solid curve), from 
above (the dashed curve), and a general truncation when each data point has its 
specific observable range (shown by the  cross for only one of the points).
{\bf Right Panel:\,} Description of the constructs used for evaluation of the 
univariate distribution from uncorrelated data truncated from below. The large 
box 
contains $N_i$ sources in the {\it comparable or associated set} of the source 
$i$ 
located in the thin rectangle defined by the dashed horizontal line.   
}
\end{figure}

\section{A BRIEF HISTORICAL REVIEW}

The process of determination of the distribution of physical characteristics 
from
truncated data has a long history starting with first and rudimentary 
observations
of stars in the disk of our galaxy.  Here I will touch upon some of the relevant
highlights.  A more detailed discussion can be found in the references cited 
below
and in a review article (Petrosian 1992).  I will limit the discussion to works
aimed at determination of the luminosity function and spatial distribution of
sources; $\Psi (L,r)$ from a magnitude limited data (one sided truncation) as
shown by the right panel in Figure 1.  Almost all the methods I will describe
ignore possible correlations between the variables and assume that these are
independent.  In the case of the luminosity function this means that $g(z)=1$, 
so
that the bivariate distribution is separable; $\Psi(L,r)=\psi(L)\rho(r)$.
Unfortunately this often unjustified simplification is prevalent even today.  
This
unnecessary assumption often can lead to erroneous conclusions.  As stressed 
below
the first task must be testing the bivariate distribution for correlations 
between
the variables.

In general, most of the methods can be divided into two categories, {\it
parametric} or {\it non parametric}.  In the former one assumes a parametric 
form
for the two functions $\psi(L)$ and $\rho(r)$.  In the non parametric methods 
one
often ends up with a description for the cumulative functions

\beq\label{cumlf}
\phi(L)=\int^\infty_L \psi(L')dL', \,\,\, \sigma (r) 
=\int^r_0(dV/dr)\rho(r')dr',
\eeq
where $V(r)$ is the volume of space (included in the observation) from the 
origin 
to distance $r$. In these relation $r$ could be any measure of distance 
including 
redshift, look back time etc.

\subsection{Parametric Methods}

The first methods were developed for investigation of the luminosity function 
and
spatial distribution of stars perpendicular to the galactic disk.  These and
subsequent applications to other sources have come to be known as correction for
{\bf Malmquist Bias} after Malmquist (1922); Eddington (1915, 1940) also 
describes
this method.  A more recent description can be found in Trumpler \& Weaver 
(1953).
These early works dealing with the distribution of {\bf stars} assume a Gaussian
distribution of absolute magnitudes ({\it i.e.}  a log-normal luminosity 
function)
and a Gaussian spatial distribution perpendicular to the disk.  Because of the
truncation there is absence of low luminosity stars at large distance so that 
the
raw observed distributions of $L$ and $r$ are biased.  The essence of the method
was to correct for this bias.  It turns out that the method used was sound but 
the
final results were incorrect because of the erroneous assumptions about the 
forms
of the distributions.  We know today that the luminosity function of stars is 
best
represented by a broken power law and that the fall of the stars perpendicular 
to
the disk is exponential and not Gaussian.  This demonstrates the shortcoming of
this and other parametric methods.  Similar parametric methods have been used 
for
extragalactic sources ({\bf galaxies and quasars}) first by Neymann \& Scott
(1959) and in numerous works ever since.

\subsection{Non parametric Methods}

One of the most commonly non parametric methods used is the so-called 
$V/V_{max}$
method first described by Kafka (1966) soon after the discovery of the quasars 
but
used most successfully by Schmidt (1968).  Independence is again assumed and the
presence or absence of density evolution is tested by a single moment, namely 
the
average value of $V/V_{max}$, where $V_{max}$ is the volume up to the maximum
redshift (or distance) that a source of luminosity $L$ can be visible given the
limiting flux $f_{min}$; $L=4\pi d_L^ 2(\Omega_i, z_{max})$ (see Figure 1).  In
the absence of evolution one expects a value of 0.5 for this average.  A more
general method was described later by Avni \& Bahcall (1980).  Of course, one 
need
not be limited only to one moment of the distribution.  More information can be
obtained by examining the distribution of $(V/V_{max})_i$ of the whole data set.

Schmidt (1968) also described a method for determination of $\phi(L)$, which was
later dubbed as the "Schmidt Estimator" by Felten (1976).  It is straightforward
to show that in absence of evolution, {\it i.e.}  for a uniform spatial
distribution, the contribution to this cumulative luminosity function of each
source is proportional to $V_{max,i}^{-1}$.  It is also easy to show that if 
there
is evolution this contribution is proportional to $\sigma_{max,i}^{-1}$ defined 
in
equation (2);  $\sigma$ was denoted as $V'$ by Schmidt.

Most other methods employ binning, which simplifies the problem conceptually 
but has several shortcomings, the primary being loss of data points in the 
incomplete bins at the truncation boundaries. Examples of these are anlysis by 
Nicole \& Segal (1978, 1983), Turner (1979) and Choloniewski (1986, 1987). As 
shown by Petrosian (1986), it turns out that all these procedures, in the limit 
of one source per bin reduce to Lynden-Bell's (1979) $C^-$ method. For a 
detailed comparison of these methods see Petrosian (1992).

\subsection{The General Method}

The right panel of Figure 1 depicts two boxes. Let us assume that number of 
data points in the narrow box is $n(L)dL$ and the number in the big box, 
excluding the narrow region is $N(L)$. It is easy to show that if the variables 
are independent, {\it i.e.} the 
luminosity function is separable in $L$ and $z$, then 

\beq\label{method}
{{n(L)dL} \over {N(L)}}  =  {{\sigma(z_{max})\psi(L)dL} \over 
{\sigma(z_{max})\phi(L)}} = - d{\rm ln} \phi(L).
\eeq
In the limiting case of one object per (narrow) bin, $n(L)=\delta(L-L_i)$ and 
$N(L)=N_i + \Theta (L-L_i)$, where $\Theta$ is the Heviside step function. 
Integration of the above equation then shows that the 
cumulative luminosity function $\phi(L)$ increases by $\delta {\rm ln} \phi(L) = 
{\rm ln} (1+N_i^{-1})$ going across the source at $L_i$. Thus, the cumulative 
luminosity function can be build in increments starting, say from the highest 
observed luminosity $L_1$, as

\beq\label{phi}
\phi(L)=\phi(L_1)\prod_{L_i>L} {\rm ln}(1+N_i^{-1}).
\eeq
The set of $N_i$ sources in the big box are  referred to as the {\it 
comparable} or the {\it associated} set of the source with luminosity $L_i$. 
It is clear that because of the complete mathematical symmetry between the two 
variables we can 
get an identical expression for the cumulative distribution $\sigma(z)$. The 
(different) associated sets are defined in a similar manner.

In the next section I will describe the extension of this method to more 
complex truncations. I will consider only bivariate distributions. The 
generalizations to multivariate distributions is straightforward.

\section{THE NEW METHODS}

The complete description of the distributions is a two step process.  The first
step is to determine whether the two variables are {\it correlated} or they are
{\it independent}, and if correlated, then find a way to account for this
correlation.  The latter step can be done only parametrically.  For the 
luminosity
function it entails the determination of the form of the luminosity evolution
function $g(z)$ such that a redefined luminosity $L_0=L/g(z)$ is uncorrelated
with, or is independent of, $z$.  The second step is to determine the univariate
distributions of the independent variables $z$ and $L_0$.

\subsection{Untruncated Data}

If $z$ and $L$ are independent then the rank $R_i$ of $z_i$ (or $L_i$) in an
untruncated sample (i.e.  a sample truncated parallel to the axes; $z>z_{min}$ 
and 
$L>L_{min}$)
will be distributed uniformly between $1$ and $N$ with an expected mean $E={1
\over 2} (N+1)$ and variance $V = {1 \over 12} (N^2 - 1)$.  We may then 
normalize
$R_i$ to have a mean of $0$ and a variance of $1$ by defining the statistic $T_i 
=
(R_i - E) / V $.  The hypothesis of independence is then rejected or accepted 
using a statistics based on the distribution of the $T_i$.  The quantity

\beq\label{tau}
\tau = { {\sum_i ( R_i - E )} \over \sqrt{\sum_i V} }
\eeq
is one choice of such a test statistic with a mean of $0$ and a variance of $1$.
The hypothesis of independence is rejected if $|\tau_{data}|$ is too large
(e.g.  $|\tau_{data}| \geq 1$ for rejection of independence at the $ 1 \ \sigma$
level).  This $\tau$ is equivalent to Kendell's $\tau$ statistic (see, {\it 
e.g.}  
Press et al. 1990)

If it turns out that $|\tau_{data}| \leq 1$ and that the variables are
independent, then the determination of the univariate distributions in each
variable is obtained ignoring the value of the other.  For example, the 
cumulative
luminosity function will be described by the histogram $\phi(L_{i+1}>L>L_i)= i$.
However, if the variables are correlated, one must then carry out a 
transformation to
remove the correlation by some parametric function, say $L_0=L/g(z)$. It is 
important to note that the transformed
data will now appear truncated in the $L_0-z$ plane with 
$L_{0,min}=L_{min}/g(z)$,
which is no longer parallel to the $z$ axis.  The use of the method described by
equation (3) is then required to obtain the univariate distributions.

\subsection{Data with One-sided Truncation}

A straightforward application of the above method to a truncated data will 
clearly give a
false correlation signal.  Efron \& Petrosian (1992), and independently Tsai
(1990), describe how this method can be applied to data with one-sided 
truncation.
The above procedure is modified as follows.  For each object define a new {\it
comparable} or {\it associated} set

\beq\label{setJ}
J_i = \{ j : L_j > L_i, \ L_j^- < L_i \},
\eeq 
where $L_j^-=4\pi d_L^2(\Omega_i, z_j)f_{min}$.  It is easy to see that this is
the same set defined above (Figure 1) as the big box containing $N_i$ sources.
(Note that the set defined in Efron \& Petrosian (1992) includes the object $i$ 
in
question.)  This is the largest subset of luminosity and volume limited data 
that
can be constructed for each point $(L_i, z_i)$.  If $z$ and $L$ are independent
then we expect the rank $R_i$ of $z_i$ (or $L_i$) in this limited set, not in 
the
whole sample to be uniformly distributed between $1$ and $N_i$.  The rest of the
procedure follows the steps described above.

Similarly the determination of the univariate distributions will require the use
of the method described in the previous section for the transformed and
uncorrelated variables.  Note again that the truncation boundary also gets
transformed in case there is a correlation between $L$ and $z$.

\subsection{Complex Truncations}

A generalization of the above method to doubly (or multiply) truncated data was
developed by Efron \& Petrosian (1999), which is valid for the most general
truncation $L_i^-<L_i<L_i^+$ and $z_i^-<z_i<z_i^+$.  The method is equivalent to
the previous method, with the associated set defined as

\beq
J_i = \{ j : L_j > L_i, \ L_i \in (L_j^-, L_j^+)\}.
\eeq
In this case, however, the distribution of the rankings (or of $\tau$)
is unknown.  If the data
are uncorrelated then $\tau$ must still have a mean of zero. But
a bootstrap method using simulations based on the  $\psi(L)$ obtained from 
the data, as described below, is required for the purpose of the estimation of 
the variance. 

For doubly (or more complexly) truncated data the comparable set is not 
completely
observed, thus a simple analytic method such as the one described in equation
(4) is not possible.  However, it turns out that a simple iterative
procedure can lead to a maximum likelihood estimate of the distributions.  Efron
and Petrosian (1999) give a thorough description of this method; for a more 
brief
and transparent description see Maloney \& Petrosian (1999).

\begin{figure}[ht]
\leavevmode\centering
\psfig{file=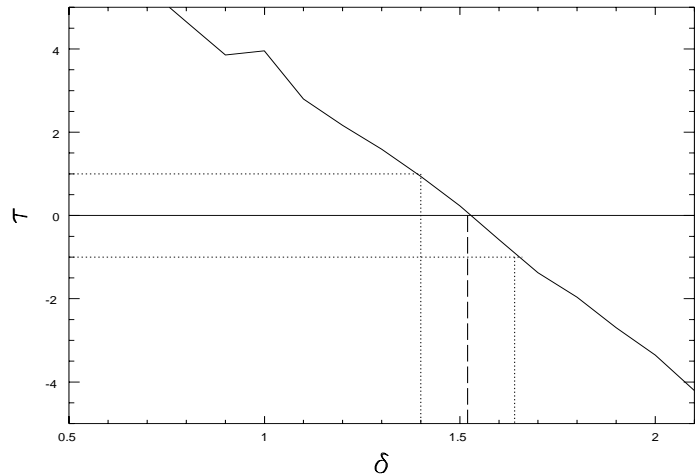,width=0.4\textwidth,height=0.6\textwidth,angle=270}
\caption{Variation of the $\tau$ statistics with exponent $\delta$ of the 
assumed 
correlation ${\bar y}\propto x^{-\delta}$. For the parent sample $\delta=1.5$.
}
\end{figure}

\subsection{Tests of the Correlation Algorithm}

Lloyd et al. (2000) describe two simulations which test how well the above 
procedures reproduce known distributions. In one simulation the method was 
applied to a single sided truncation  of an uncorrelated bivariate  parent
distribution. The rank test applied to the untruncated simulated points gave a 
value of $\tau=0.9$ But when applied to the truncated data without considering 
the effects of the truncation resulted in $\tau > 5.0$, indicating the presence 
of 
a strong 
(of course false) correlation that is introduced by the truncation. However, 
when the test 
was 
carried out correctly by accounting for  the effects of the truncation (as 
described above) it was found that $\tau=0.6$, recovering the fact that the 
variables were 
independent. In a second set the simulated parent sample had a strong 
anti correlation, with variables $y$ and $x$ correlated with average value of 
${\bar y}\propto x^{-1.5}$. Application of the correlation to the untruncated 
randomly selected sample, as expected, gave a negative value for the statistic;  
$\tau<-5$. 
The data was truncated  and the method applied blindly. This resulted in the 
value 
of $\tau=+2.5$, or a (false) positive correlation. However, when applied 
properly 
it 
gave a value of $\tau=-4.0$ indicating the presence of the anti correlation at 4 
sigma level. Assuming a correlation of ${\bar y}\propto x^{-\delta}$, the data 
was transformed accordingly and the method applied to the new data. Figure 2 
shows the variation of $\tau$ with the exponent $\delta$, from which we deduce a 
mean value 
and one sigma range of $\delta = 1.51\pm 0.11$. This is a strong support for 
the accuracy of the method.

Further simulations are required to test the methods for determination of the 
distribution and correlations for more complex truncations.

\section{SOME RESULTS FROM AGN DATA}

Our application of these methods to a combine set of several optically selected 
samples 
of quasars are described in Maloney \& Petrosian (1999).  Here is a brief 
summary.

$\bullet$ We found a strong correlation between luminosity and redshift,
indicating the presence of a rapid luminosity evolution.

$\bullet$ The parametric model of luminosity evolution $(1+z)^{k'}$ provides a
better description of the data than the model $e^{k t(z)}$, where $t(z)$ is the
look back time.  Neither parameterization perfectly removes the correlation in 
all
areas of the $L - z$ plane.  In order to better model this evolution future
analyses of quasar evolution could consider parametric forms, with more than one
free parameter.  Some of the more complex forms of evolution suggested in the
literature, {\it e.g.}  the used by La Franca \& Cristiani (1996) are equivalent
with a luminosity evolution form that contains two independent free parameters.

Given the form of the luminosity evolution we make the simple transformation of
all luminosities to their hypothetical present epoch values, $L_0=L/g(z)$, so 
that
$L_0$ and $z$ are uncorrelated.  This allows us to use our methods to determine
the univariate distributions of $z$ (the density evolution) and $L_0$ (the local
luminosity function).

$\bullet$ We find that the co-moving density of quasars also evolves, but its
extent depends on the cosmological model.  For example, for the Einstein$-$de
Sitter model $\rho \sim (1 + z)^{2.5}$ for low redshifts and rapidly declines as
$\rho \sim (1 + z)^{-5}$ for $z > 2$.  This is much slower evolution than is
obtained when one (incorrectly) assumes a pure density evolution model; $g(z)=1$
(Schmidt 1968; Miyaji, Hasionger \& Schmidt 1998).

$\bullet$ The cumulative local luminosity function $\phi(L_o)$ has the double
power law form found previously (Caditz \& Petrosian 1990, La Franca \& 
Cristiani
1996), with a break luminosity of $L_\ast = 6 \times 10^{29}$ erg / (sec Hz), in
the Einstein$-$de Sitter model.  The power-law indices at the low and high 
luminosity ends are -1.5 and -2.3 in rough agreement with previous estimates. 
There is however, some evidence for evolution of the shape of the luminosity 
function. More data is required for a quantitative description of this 
evolution.


\end{document}